\documentclass[prc,a4paper,twocolumn,showpacs,superscriptaddress,showkeywords]{revtex4-1}
\usepackage[dvips]{graphics,graphicx,geometry,epsfig}
\usepackage{amsmath}
\usepackage{wasysym}
\usepackage{bm}
\usepackage{natbib}
\usepackage{dcolumn}
\usepackage{lineno}

\begin{document}

\title{Precision measurements of the $^{60}$Co $\beta$-asymmetry parameter in search for tensor currents in weak interactions}

\author{F. Wauters}
\email{Frederik.Wauters@fys.kuleuven.be}
\affiliation{K. U. Leuven, Instituut voor Kern- en Stralingsfysica, Celestijnenlaan 200D, B-3001 Leuven, Belgium}
\author{I. Kraev}
\affiliation{K. U. Leuven, Instituut voor Kern- en Stralingsfysica, Celestijnenlaan 200D, B-3001 Leuven, Belgium}
\author{D. Z\'akouck\'y}
\affiliation{Nuclear Physics Institute, ASCR, 250 68 \v{R}e\v{z}, Czech Republic}
\author{M. Beck}
\altaffiliation[Present address: ]{Westf\"alische Wilhelms-Universit\"at M\"unster, Institut f\"ur Kernphysik, Wilhelm-Klemm-Stra\ss e 9, 48149 M\"unster, Germany}
\affiliation{K. U. Leuven, Instituut voor Kern- en Stralingsfysica, Celestijnenlaan 200D, B-3001 Leuven, Belgium}
\author{M. Breitenfeldt}
\affiliation{K. U. Leuven, Instituut voor Kern- en Stralingsfysica, Celestijnenlaan 200D, B-3001 Leuven, Belgium}
\author{V. De Leebeeck}
\affiliation{K. U. Leuven, Instituut voor Kern- en Stralingsfysica, Celestijnenlaan 200D, B-3001 Leuven, Belgium}
\author{V. V. Golovko}
\altaffiliation[Present address: ]{Queen's Particle Astrophysics, Queen's University, Kingston, Ontario K7L 3N6, Canada}
\affiliation{K. U. Leuven, Instituut voor Kern- en Stralingsfysica, Celestijnenlaan 200D, B-3001 Leuven, Belgium}
\author{V. Yu. Kozlov}
\affiliation{K. U. Leuven, Instituut voor Kern- en Stralingsfysica, Celestijnenlaan 200D, B-3001 Leuven, Belgium}
\author{T. Phalet}
\affiliation{K. U. Leuven, Instituut voor Kern- en Stralingsfysica, Celestijnenlaan 200D, B-3001 Leuven, Belgium}
\author{S. Roccia}
\affiliation{K. U. Leuven, Instituut voor Kern- en Stralingsfysica, Celestijnenlaan 200D, B-3001 Leuven, Belgium}
\author{G. Soti}
\affiliation{K. U. Leuven, Instituut voor Kern- en Stralingsfysica, Celestijnenlaan 200D, B-3001 Leuven, Belgium}
\author{M. Tandecki}
\affiliation{K. U. Leuven, Instituut voor Kern- en Stralingsfysica, Celestijnenlaan 200D, B-3001 Leuven, Belgium}
\author{I. S. Towner}
\affiliation{Cyclotron Institute, Texas A \& M University, College Station, Texas 77845, U.S.A.}
\author{E. Traykov}
\affiliation{K. U. Leuven, Instituut voor Kern- en Stralingsfysica, Celestijnenlaan 200D, B-3001 Leuven, Belgium}
\author{S. Van Gorp}
\affiliation{K. U. Leuven, Instituut voor Kern- en Stralingsfysica, Celestijnenlaan 200D, B-3001 Leuven, Belgium}
\author{N. Severijns}
\affiliation{K. U. Leuven, Instituut voor Kern- en Stralingsfysica, Celestijnenlaan 200D, B-3001 Leuven, Belgium}

\date{\today}

\begin{abstract}
The $\beta$-asymmetry parameter $\widetilde{A}$ for the Gamow-Teller decay of $^{60}$Co was measured by polarizing the radioactive nuclei with the brute force low-temperature nuclear-orientation method. The $^{60}$Co activity was cooled down to milliKelvin temperatures in a $^3$He-$^4$He dilution refrigerator in an external 13 T magnetic field. The $\beta$~ particles were observed by a 500 ${\mu}m$ thick Si PIN diode operating at a temperature of about 10 K in a magnetic field of 0.6 T. Extensive GEANT4 Monte-Carlo simulations were performed to gain control over the systematic effects. Our result, $\widetilde{A} = -1.014(12)_{stat}(16)_{syst}$, is in agreement with the Standard-Model value of $-0.987(9)$, which includes recoil-order corrections that were addressed for the first time for this isotope.  Further, it enables limits to be placed on possible tensor-type charged weak currents as well as other physics beyond the Standard Model.
\end{abstract}




\maketitle

\section{Introduction}

In the Standard Electroweak Model the weak interaction has a vector -- axial-vector structure ($V-A$) implying absence of tensor ($T$), scalar ($S$), and pseudoscalar ($P$) components. In nuclear $\beta$~decay the contribution of the pseudoscalar component can be excluded because of the non-relativistic behavior of nucleons. This argument, however, does not apply to the scalar and tensor components.  Currently, there is no theoretical motivation for their absence \cite{Lee1956}. Present experimental limits from neutron and nuclear $\beta$~decay restrict their potential contribution to about $8~\%$ in the amplitudes \cite{Severijns2006} and experimental efforts to improve these constraints are ongoing in both nuclear $\beta$ beta decay \cite{Adelberger:99}-\cite{Kozlov:2008} and free neutron decay \cite{pattie09}-\cite{kozela09}.

A measurement of the angular distribution of the $\beta$~radiation from oriented nuclei is potentially very sensitive to deviations from the Standard-Model weak interaction. This angular distribution is given by: \cite{jackson57}
\begin{eqnarray}
\label{eqn:jtw}  W(\theta) \propto \left[ 1 + b \frac{m}{E_e} +
\frac{\bf{p}}{E_e} \cdot A \bf{J}  \right] ,
\end{eqnarray}
\noindent with $E_e$ and $\bf{p_e}$ the total energy and momentum
of the $\beta$~particle, $m$ the rest mass of the electron, $\textbf{J}$ the nuclear vector polarization and $b$ the Fierz interference term.

The experimental observable $\widetilde{A}$ for an allowed pure Gamow-Teller decay can be written as \cite{Severijns2006}:
\begin{eqnarray}
\label{eq:asym}
\widetilde{A}_{GT}^{\beta^\mp}   & \equiv & \frac{A}{1 \pm \frac{m}{E_e}b}
\nonumber \\
& \simeq & A_{SM} + \lambda \Big{[} \frac{ \alpha Z m } {p_e} \Im \left( \frac{ C_T + C^{\prime}_T }{ C_A } \right)
\nonumber \\ & & ~~~~
 + \frac{\gamma m} {E_e} \Re \left(\frac{ C_T + C^{\prime}_T }{ C_A } \right)
\nonumber \\
& & ~~~~ \pm \Re \left( \frac{ C_T C^{\prime *}_T } { C_A^2 } \right)
  \pm \frac{ |C_T|^2 + |C^{\prime}_T|^2 }{ 2 C_A^2 }\Big{]}
\nonumber \\
& \simeq & A_{SM} + \lambda \frac{\gamma m}{E_e} \Re \left( \frac{ C_T + C^{\prime}_T}{C_A} \right) ,
\end{eqnarray}
\noindent with $C_T$, $C'_T$ and $C_A$ coupling constants of the tensor and axial-vector parts of the weak interaction Hamiltonian, as introduced by Jackson, Treiman and Wyld \cite{jackson57}. Primed (unprimed) coupling constants are for the parity conserving (violating) parts of the interactions, respectively (maximum parity violation was assumed for the axial-vector part of the interaction). Further, the upper (lower) sign refers to $\beta^{-}$($\beta^{+}$) decay,  and $\gamma = [ 1 - \left( {\alpha Z} \right)^2 ]^{1/2}$ with $\alpha$ the fine structure constant and $Z$ the atomic number of the daughter isotope. Also, $\lambda=1$ for $J \to J- 1$, $\lambda=1/(J+1)$ for $J \to J$ and $\lambda  = {{ - J} \mathord{\left/ {\vphantom {{ - J} {\left( {J + 1} \right)}}} \right. \kern-\nulldelimiterspace} {\left( {J + 1} \right)}} $ for $J \to J+1$ transitions.

The Standard-Model prediction for the $\beta$-asymmetry parameter is $A_{SM} = \mp \lambda$ + \emph{recoil corrections}. The recoil corrections are determined by the induced form factors \cite{Holstein1974} and will be addressed in section IV.

To obtain the last line of Eq.~(\ref{eq:asym}) we make two assumptions.  First, existing limits on the imaginary term, $\Im (C_T + C^{\prime}_T)/C_A$, are already at the $1~\%$ level \cite{Huber2003} suggesting this term is negligible.
Second, the couplings $C_T/C_A$ and $C^{\prime}_T/C_A$ are presumably small in order that second-order terms such as $|C_T|^2/|C_A|^2$ can be neglected.  Then it is evident that any departure in the measured value $\widetilde{A}$ from the
Standard-Model value $A_{SM}$ is sensitive to the tensor couplings $(C_T + C^{\prime}_T)$.
Further, as the factor $\gamma$~is always of order unity, the sensitivity to these tensor couplings can be enhanced by selecting a $\beta$~decay with a low endpoint energy.

A powerful technique for this kind of measurement is the Low Temperature Nuclear Orientation method (LTNO) \cite{Stone:1986}. An excellent illustration of the potential of this method is the well-known experiment performed in 1957 with $^{60}$Co by C. S. Wu {\it et al.}, which established the violation of parity in weak interactions \cite{Wu:1957}.  This experiment was later repeated with better precision by Chirovsky {\it et al.} \cite{Chirovsky1984}.

Table~\ref{tab:Abig} summarizes the most precise $\beta$-asymmetry measurements for superallowed mixed Fermi/Gamow-Teller and pure Gamow-Teller (GT) nuclear $\beta$~decays. For pure Fermi transitions $\widetilde{A} \equiv 0$. Allowed $J^\pi \rightarrow J^\pi$ transitions between non-analog states ({\it i.e.} with different isospin), which are in principle of pure GT type according to weak interaction selection rules, are not considered here as they can contain a small Fermi component from isospin mixing caused  by the electromagnetic interaction (see {\it e.g.}~Ref.~\cite{Raman1975,Groves1982,Lee1983,Lee1985,Schuurmans2000,Severijns2005}), rendering them uninteresting for weak interaction studies.


\begin{table*}[!htb]
\caption{Overview of the measurements of the $\beta$-asymmetry parameter $A$ for pure GT transitions and isospin $T$ = 1/2 mixed F/GT mirror transitions. All results were obtained with the LTNO method, except for the last four results. $^{19}$Ne was polarized with the help of a Stern-Gerlach magnet, $^{29}$P and $^{35}$Ar were produced and polarized by polarization transfer reactions. Neutron decay results are not included here.}
\begin{center}
\begin{tabular*}{0.9\textwidth}{@{\extracolsep{\fill}}lccc}
  \hline
  \hline
& & \\
Isotope& Transition & $A$ & $A_{SM}^{(a)}$\smallskip\\
\hline
& & \\
$^{60}$Co& $5^ +  \xrightarrow{{GT}}4^ +$ & $-1.01\left(2\right)$$^{\left( 1 \right)}$ & $-1$ \\
&& $-0.972\left(34\right)$$^{\left( 2 \right)}$ & \smallskip \\
$^{114}$In& $1^ +  \xrightarrow{{GT}}0^ +$ & $-0.990\left(14\right)$$^{\left( 3 \right)}$  & $-1$ \smallskip \smallskip \\
$^{127}$Te$^{\text{g}}$& $ {{3 \mathord{\left/ {\vphantom {3 2}} \right.
 \kern-\nulldelimiterspace} 2}} ^ +  \xrightarrow{{GT}} {{5 \mathord{\left/
 {\vphantom {5 2}} \right.
 \kern-\nulldelimiterspace} 2}} ^ +$ & $0.569\left(51\right)$$^{\left( 4 \right)}$  & 0.6 \smallskip \smallskip \\

$^{129}$Te$^{\text{g}}$& $ {{3 \mathord{\left/ {\vphantom {3 2}} \right.
 \kern-\nulldelimiterspace} 2}} ^ +  \xrightarrow{{GT}} {{5 \mathord{\left/
 {\vphantom {5 2}} \right.
 \kern-\nulldelimiterspace} 2}} ^ +$ & $0.645\left(59\right)$$^{\left( 4 \right)}$  & 0.6 \smallskip  \smallskip\\

 $^{133}$Xe$^{\text{g}}$& $ {{3 \mathord{\left/ {\vphantom {3 2}} \right.
 \kern-\nulldelimiterspace} 2}} ^ +  \xrightarrow{{GT}} {{5 \mathord{\left/
 {\vphantom {5 2}} \right.
 \kern-\nulldelimiterspace} 2}} ^ +$ & $0.598\left(73\right)$$^{\left( 4 \right)}$ & 0.6 \smallskip \\


  $^{17}$F&${{5 \mathord{\left/
 {\vphantom {5 2}} \right.
 \kern-\nulldelimiterspace} 2}} ^ +  \to  {{5 \mathord{\left/
 {\vphantom {5 2}} \right.
 \kern-\nulldelimiterspace} 2}} ^ +$  & $0.960\left(82\right)$$^{\left( 5 \right)}$ & 0.99739(18)\\
&$T={\raise0.5ex\hbox{$\scriptstyle 1$}
\kern-0.1em/\kern-0.15em
\lower0.25ex\hbox{$\scriptstyle 2$}}$ mirror& \smallskip\smallskip\\

$^{19}$Ne&${{1 \mathord{\left/
 {\vphantom {1 2}} \right.
 \kern-\nulldelimiterspace} 2}} ^ +  \to  {{1 \mathord{\left/
 {\vphantom {1 2}} \right.
 \kern-\nulldelimiterspace} 2}} ^ +$  & $-0.0391\left(14\right)$$^{\left( 6 \right)}$ & $-0.04166(95)$\\
&$T={\raise0.5ex\hbox{$\scriptstyle 1$}
\kern-0.1em/\kern-0.15em
\lower0.25ex\hbox{$\scriptstyle 2$}}$ mirror& \smallskip\smallskip\\

$^{29}$P&${{1 \mathord{\left/
 {\vphantom {5 2}} \right.
 \kern-\nulldelimiterspace} 2}} ^ +  \to  {{1 \mathord{\left/
 {\vphantom {5 2}} \right.
 \kern-\nulldelimiterspace} 2}} ^ +$  & $0.681\left(86\right)$$^{\left( 7 \right)}$ & 0.6154(46)\\
&$T={\raise0.5ex\hbox{$\scriptstyle 1$}
\kern-0.1em/\kern-0.15em
\lower0.25ex\hbox{$\scriptstyle 2$}}$ mirror& \smallskip\smallskip\\

$^{35}$Ar&$ {{3 \mathord{\left/
 {\vphantom {1 2}} \right.
 \kern-\nulldelimiterspace} 2}} ^ +  \to  {{3 \mathord{\left/
 {\vphantom {1 2}} \right.
 \kern-\nulldelimiterspace} 2}} ^ +$  & $0.49\left(10\right)$$^{\left( 8 \right)}$ & 0.4371(36)\\
&$T={\raise0.5ex\hbox{$\scriptstyle 1$}
\kern-0.1em/\kern-0.15em
\lower0.25ex\hbox{$\scriptstyle 2$}}$ mirror&\\

$^{35}$Ar&$ {{3 \mathord{\left/
 {\vphantom {1 2}} \right.
 \kern-\nulldelimiterspace} 2}} ^ +  \to  {{3 \mathord{\left/
 {\vphantom {1 2}} \right.
 \kern-\nulldelimiterspace} 2}} ^ +$  & $0.427\left(23\right)$$^{\left( 9 \right)}$ & 0.4371(36)\\
&$T={\raise0.5ex\hbox{$\scriptstyle 1$}
\kern-0.1em/\kern-0.15em
\lower0.25ex\hbox{$\scriptstyle 2$}}$ mirror&\\

\smallskip \\
\hline
\hline
\end{tabular*}\\
\begin{flushleft}
\scriptsize $^{\left( 1 \right)}$Ref. \cite{Chirovsky1984}; $^{\left( 2 \right)}$Ref. \cite{Hung1976}; $^{\left( 3 \right)}$Ref. \cite{WautersIn2009}; $^{\left( 4 \right)}$Ref.  \cite{Vanneste1986}; $^{\left( 5 \right)}$Ref. \cite{Severijns1988,Severijns1989}; $^{\left( 6 \right)}$Ref. \cite{Calaprice1975}; $^{\left( 7\right)}$Ref. \cite{Masson1990}; $^{\left( 7\right)}$Ref. \cite{Garnett1988}; $^{\left( 7\right)}$Ref. \cite{Converse1993}; $^{(a)}$ Neglecting recoil corrections (see Sect. IV); values for the T=1/2 mirror transitions taken from \cite{Severijns2008}
\end{flushleft}
\end{center}
\label{tab:Abig}

\end{table*}

For the most precise measurements to date with $^{60}$Co \cite{Chirovsky1984} and $^{114}$In \cite{WautersIn2009}, accuracies of $2~\%$ and $1.5~\%$ have been obtained. To improve significantly on the existing constraints for a tensor-type weak interaction the $\beta$-asymmetry parameter has to be determined with a precision of at least this order or even better, depending on the isotope in question and its $\beta$~endpoint energy, see Eq.~(\ref{eq:asym}).

For this purpose a new Brute Force LTNO setup was installed at the Katholieke Universiteit Leuven. The first tests of the setup were performed in 2005 and were reported in Ref.~\cite{Kraev2005}. Here we report on a measurement of the $\beta$-asymmetry parameter for the $5^+\rightarrow4^+$ pure Gamow-Teller $\beta^-$ transition in the decay of $^{60}$Co ($t_{1/2} = 1925.28(14)$ d), with an endpoint energy of 317.9~keV and $\log ft$~=~7.512(2) \cite{NNDC}.

\section{The experiment}

\subsection{Experimental setup}
\begin{figure}[ht]
\centering
\includegraphics[scale=0.5]{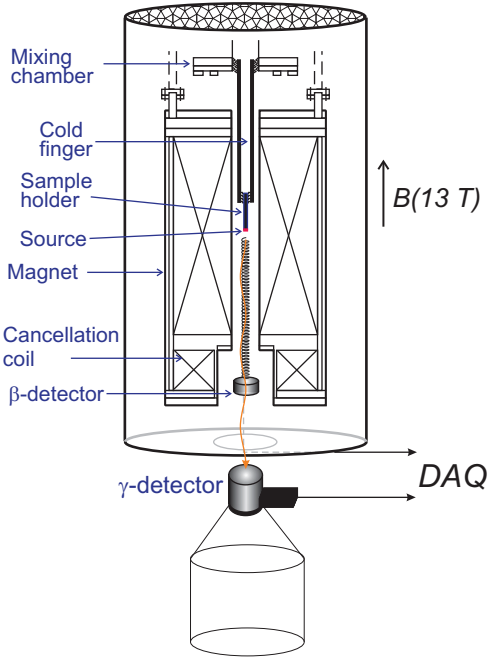}
\caption {Schematic view of the lower part of the experimental
setup.} \label{fig:sketch}
\end{figure}

The experimental setup was described in detail in Ref.~\cite{Kraev2005}. The $^{60}$Co nuclei were oriented with the Brute Force LTNO method \cite{Stone_Bible:Ch9}, in which the nuclei are embedded in a non-magnetic host foil that is cooled to milliKelvin temperatures while being exposed to a high external magnetic field. The experimental setup consisted of a $^3$He-$^4$He dilution refrigerator equipped with a superconducting magnet. A schematic view of the bottom part of the setup is shown in Fig.~\ref{fig:sketch}. The magnet has an internal diameter of 36 mm, with a special inset at the bottom of the magnet to house the particle detector. The choice of a particle detector is determined by its ability to work both at temperatures close to liquid He temperature and in a strong magnetic field, which may reach 1 T at the site of the detector. A Si PIN photodiode from Hamamatsu Photonics with a thickness of 500 $\mu$m and a surface area of 9$\times$9 mm$^{2}$ has been tested and showed good behavior under such conditions \cite{Wauters2009,Kraev2005}. A Si detector of this thickness is able to fully stop electrons with energies up to $\sim$350~keV, while at the same time having a rather low sensitivity to $\gamma$~radiation in comparison to a thicker Si detector or a high-purity Ge detector. Another advantage of using a Si detector is that the electron backscattering probability for Si is significantly lower than for Ge (by a factor of about 2.3 \cite{Tabata1971}).
\\
The sample for the Brute Force LTNO measurement was produced by diffusing $^{60}$Co and $^{57}$Co activity into a $99.99$+~\% pure 20 $\mu$m thin Cu foil.
Copper is a suitable host material since it has good thermal conductivity, is easy to solder and has a rather low $Z$, which is important to minimize scattering of $\beta$~particles in the foil. Further, copper is a non-ferromagnetic material such that no internal hyperfine field is present. The Co nuclei thus only feels the external magnetic field, slightly altered by a Knight shift of $+5.2(2)~\%$ \cite{Stone_Bible:Ch9,Wada1971}. The source was prepared by first heating the Cu foil with the dried Co activity under a hydrogen atmosphere to 600 $^\circ$C to anneal the foil. Then the temperature was increased up to 850 $^\circ$C and maintained for 5 minutes for the actual diffusion of the Co activities into the foil. Thereafter, the temperature was decreased down to 500 $^\circ$C to anneal the foil once more for about 24 hours in order to remove any possible remaining oxidation. The diffusion coefficient of Co in Cu is known \cite{Mackliet1958} so that the depth profile of the cobalt ions can be calculated \cite{Kraev2006}. From this, we learned that about $90~\%$ of the activity was sitting in a 3~$\mu$m thick layer below the surface.

\begin{figure}[ht]
\centering
\includegraphics[width = 0.8\columnwidth]{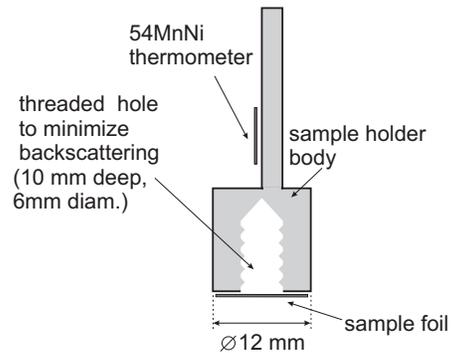}
\caption {Bottom part of the sample holder. The sample is soldered on to the bottom perpendicular to the magnetic field direction.} \label{fig:SampleHolder}
\end{figure}

The foil was soldered on to the sample holder, which is, via the cold finger, in thermal contact with the $^{3}$He-$^{4}$He mixing chamber of the dilution refrigerator, where the lowest temperature is reached. To avoid a temperature gradient over the sample holder and to maximize its temperature conductivity it is made from oxygen-free copper. The bottom part has a 10 mm deep, 6 mm wide threaded hole behind the activity spot to reduce backscattering of electrons (see Fig.~\ref{fig:SampleHolder}). Geometrically, the foil is positioned in the center of the magnetic field, perpendicular to the vertical magnetic field direction.
\\
In order to determine temperatures in the mK region a nuclear orientation thermometer $^{54}$Mn(Ni) was soldered on to the sample holder as well. Since all nuclear and hyperfine properties of this sample are accurately known and the degree of orientation in this kind of experiment follows a Boltzmann distribution, there is a one to one relation between the anisotropy of the 835~keV $\gamma$~line from the decay of $^{54}$Mn and the temperature of the sample holder \cite{Stone_Bible:16}. \\
The purpose of the $^{57}$Co in the copper sample foil itself, in addition to the $^{60}$Co, was to cross check the degree of orientation of the $^{60}$Co nuclei.

To register $\gamma$~rays, a high-purity Ge detector was installed outside the refrigerator at a distance of about 30 cm from the center of the magnetic field (Fig.~\ref{fig:sketch}). To minimize the effect of the magnetic field on this detector a cancelation coil was included in the bottom part of the magnet such that the field strength at the position of this detector is only about $2~\%$ of the central value.

\subsection{Data taking}

\begin{figure}[ht]
\centering
\includegraphics[width = 1.05\columnwidth]{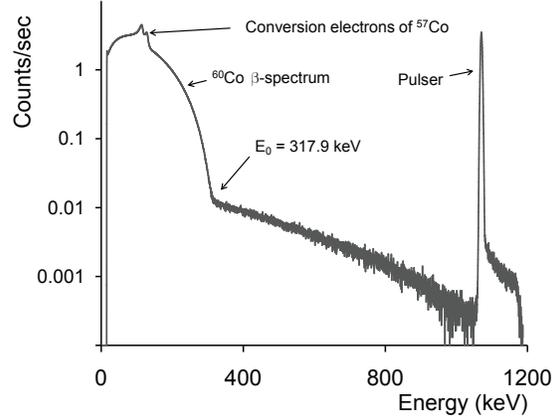}
\caption {$\beta$~Spectrum observed with the Si particle detector operating at a temperature of about 10 K and in a magnetic field of 0.6T. $E_0$ indicates the $\beta$ spectrum endpoint energy.} \label{fig:ExpBetaSpectrum}
\end{figure}

A $\beta$~spectrum measured with the Si PIN diode detector is shown in Fig.~\ref{fig:ExpBetaSpectrum}. The background seen above the $^{60}$Co endpoint at 317.9~keV is mostly due to Compton scattering of the 1173~keV and 1332~keV $\gamma$~lines in the decay $^{60}$Co. The contribution from the very weak $\beta$~decay branch of $^{60}$Co with an endpoint energy of 1491~keV and from the $^{54}$Mn 835~keV $\gamma$~line to this Compton tail is negligible. The peak at the end of the spectrum comes from the pulse generator. The observed pulser count rate is used for dead-time correction. The energy calibration was obtained by inserting a $^{207}$Bi conversion electron source in the system before and after the measurement. From this the energy resolution could be determined to be 6~keV at 500~keV

\begin{figure}[ht]
\centering
\includegraphics[width = 1.05\columnwidth]{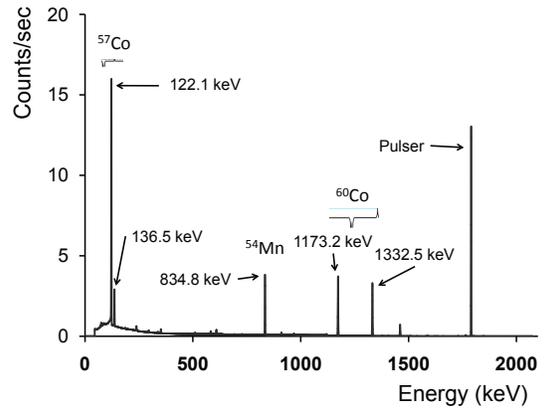}
\caption {$\gamma$ Spectrum observed with the HPGe detector. The energies of the main $\gamma$~lines are indicated.} \label{fig:ExpGammaSpectrum}
\end{figure}

The spectrum observed by the HPGe $\gamma$~detector is shown in Fig.~\ref{fig:ExpGammaSpectrum}. The 835~keV E2 $\gamma$~ transition in the decay of $^{54}$Mn was used to determine the temperature of the sample in the milliKelvin region. The $\gamma$~lines from $^{57}$Co and $^{60}$Co were used to determine the degree of nuclear orientation for the Co nuclei. The spectrum is seen to be very clean, apart from some natural background radiation lines. The FWHM of the full energy peaks was found to be 2.2~keV at 1.33 MeV.

The experimental angular distribution of $\beta$~particles emitted by polarized nuclei for an allowed $\beta$~decay can be written as
\begin{equation}
W\left( \theta  \right) = 1 + f \frac{v}{c}\tilde{A} P Q_1 \cos \theta \label{eq:W} ,
\end{equation}
where $\theta$ is the emission angle of the $\beta$~particle with respect to the polarization axis, $v/c$ is the $\beta$~ particle initial velocity relative to the speed of light, $f$ is the fraction of the nuclei that feel the full orienting interaction $\mu B$ (with $\mu$ the nuclear magnetic moment of the mother isotope and $B$ the total magnetic field the nuclei feel), $P$ is the degree of nuclear polarization, $Q_1$ takes into account the solid angle as well as effects of the magnetic field and scattering, and $\tilde{A}$ is the $\beta$-asymmetry parameter given in Eq.~(\ref{eq:asym}).

Experimentally, the angular distribution $W\left( \theta\right)$, Eq.~(\ref{eq:W}), is obtained as the ratio $ {{N\left( \theta\right)_{cold} } \mathord{\left/ {\vphantom {{N\left( \theta\right)_{cold} } {N\left( \theta  \right)_{warm} }}} \right. \kern-\nulldelimiterspace} {N\left( \theta  \right)_{warm} }} $, with $N\left( \theta  \right)_{cold}$ the count rate of $\beta$~particles at low temperature ({\it i.e.} below 100 mK, ``cold", polarized nuclei) and $N\left( \theta\right)_{warm}$ the count rate when the emission is isotropic ({\it i.e.} at $\ge$ 1 K, ``warm", unpolarized nuclei). In the present configuration the angular distribution function can be measured at two fixed angles with respect to the magnetic field direction, {\it i.e.} 0$^\circ$ and 180$^\circ$.

\begin{figure}[ht]
\centering
\includegraphics[width = 0.98\columnwidth]{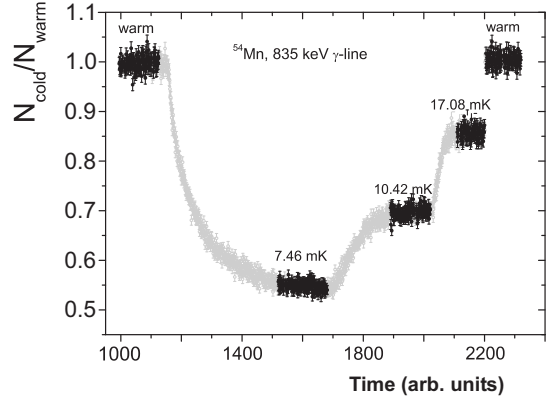}
\caption {Normalized count rate of the 835~keV $\gamma$~line of the $^{54}$Mn(Ni) thermometer. The black areas are the stable temperature regions that were used for analysis.} \label{fig:Mn54CoolingCurve}
\end{figure}

The measurement consisted of two temperature cycles. During the first, a 13 T external field was applied and after first taking $warm$ reference data the sample was cooled down to the lowest temperature, {\it i.e.} 7.46(6) mK. Thereafter, the temperature was raised to 10.4(1)~mK and 17.1(1)~mK after which, as a cross check, $warm$ data were again taken. The second temperature cycle was performed with a 9 T external field, cooling to the lowest attainable temperature, 7.38(6) mK. Also in this case isotropic, $warm$ data were taken before and after the cooldown. Fig.~\ref{fig:Mn54CoolingCurve} shows the normalized count rate of the 835~keV $\gamma$~line of the $^{54}$Mn thermometer during the 13 T measurement cycle, from which the temperature of the sample was deduced. Each data point represents 600~s of counting. Data blocks, which were used for analysis, are indicated in black.

\section{Analysis}

To determine the experimental $\beta$~anisotropy only the upper half of the $\beta$~spectrum was considered as the lower part was too much distorted by scattered electrons and was, in addition, contaminated with the conversion electron lines from the decay of $^{57}$Co.

Before $\widetilde{A}$ can be extracted from Eq.~(\ref{eq:W}), all other factors have to be determined with sufficient precision. The nuclear polarization $P$ depends on the temperature of the sample and on the magnetic hyperfine interaction $\mu B$ where $\mu(^{60}$Co$) = +3.790(8)\mu_{\text{N}}$ \cite{Stone2005} and $B = B_{ext}(1+K)$ with the Knight Shift for Co in Cu being $K$ = +0.052(2) \cite{Stone_Bible:Ch9,Wada1971}. The fraction $f$ was obtained from the $\gamma$~anisotropies of $^{57}$Co and $^{60}$Co after the temperature had been first obtained from the $^{54}$Mn(Ni) thermometer. The factor $v/c$ requires the knowledge of the initial velocity of the $\beta$~particle. The solid-angle correction factor $Q_1$, as calculated in the past, took into account the finite source and detector sizes and the detector efficiencies.  Further, it assumed a straight pathway between source and detector. This approach is clearly not valid any more when one takes into account the fact that the electrons can (back)scatter in the source or on the detector and that in this specific experiment their trajectories are also heavily influenced by the strong magnetic field. A GEANT4 \cite{geant:2003} based Monte-Carlo routine was developed to deal with these effects \cite{WautersGeant2009}. For the final analysis the complete experiment was therefore simulated assuming for $\widetilde{A}$ the Standard-Model value and using for the fraction $f$ and the nuclear polarization $P$ the values obtained from the analysis of the $^{57}$Co, $^{60}$Co, and $^{54}$Mn $\gamma$~ray anisotropies. The experimental $\beta$-asymmetry parameter was then extracted from the ratio
\begin{equation}
\frac{\left[W\left( \theta  \right) - 1\right]_{experiment}}{\left[W\left(\theta\right)-1\right]_{simulation}} = \frac{\widetilde{A}_{experiment}}{\widetilde{A}_{Standard \, Model}} \label{eq:Extract A} ,
\end{equation}
assuming that the simulations deal correctly with the coefficients $v/c$ and $Q_1 \cos \theta$.

Below, these different steps in the analysis to obtain $\widetilde{A}$ are discussed in more detail.

\subsection{Experimental anisotropies}


Unlike conventional LTNO experiments, Brute Force LTNO experiments might set the fraction $f$ to be $f=1$ since, with the use
of only an external magnetic field, one might expect all nuclei to feel the full polarizing hyperfine interaction $\mu B$. Nevertheless, the anisotropies of the $\gamma$~lines from the $^{57}$Co and the $^{60}$Co nuclei showed an attenuation of about $7~\%$ with respect to the values expected on the basis of the respective hyperfine interaction strengths $\mu B$, independent of the temperature and of the external magnetic field strength. The exact value of this attenuation was subsequently determined from the anisotropies of the 122~keV and 136~keV $\gamma$~rays of $^{57}$Co and the 1173~keV and the 1332~keV $\gamma$~rays of $^{60}$Co leading to a fraction $f_{\rm Co} = 0.928 (4)$, which was then used in the analysis of the $\beta$~anisotropies. A similar attenuation of the anisotropy with respect to the expected value was previously reported in other Brute Force experiments as well \cite{Stone_Bible:Ch9,Hutchison1992,Nuytten1982}. This reduction of the anisotropy is explained as a chemical effect of insolubility, internal oxidation or impurity clustering, depending on the sample preparation procedures and the host-impurity combination \cite{Stone_Bible:Ch9,Hutchison1992}.

\begin{figure}[ht]
\centering
\includegraphics[width = 1.00\columnwidth]{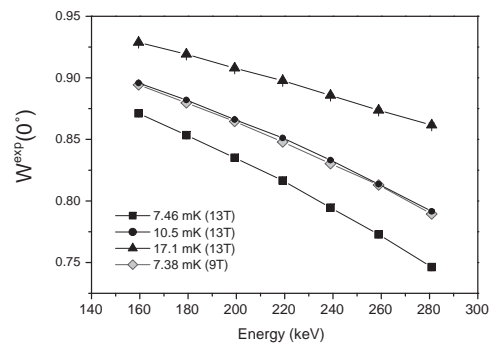}
\caption {Experimental $\beta$~anisotropies as a function of energy. The error bars are smaller than the size of the symbols. The last bin (from 290~keV to 320~keV) was omitted as the count rate in there originated mainly from Compton-scattered $\gamma$ rays (see Sec.~III.B).} \label{fig:ExpBetaAnisotropies}
\end{figure}

The $\beta$~spectrum between 150~keV and 320~keV (Fig.~\ref{fig:ExpBetaSpectrum}) was divided into eight bins. The anisotropies for the four different temperature points taken at 13 T and 9 T are shown as a function of energy in Fig.~\ref{fig:ExpBetaAnisotropies}. The decrease of anisotropy towards lower energies is due to the factor $v/c$, Eq.~(\ref{eq:W}), as well as to an increase of scattered events. Note that, as a consequence of the applied high magnetic field, the detector has an effective solid angle close to $2\pi$. To obtain reliable results from the raw data presented in Fig.~\ref{fig:ExpBetaAnisotropies}, all disturbing and usually energy-dependent effects have to be understood and correctly taken into account. This was done with the GEANT4 based Monte-Carlo simulation code.

\subsection{Monte-Carlo simulation of the experiment}

\begin{figure}[ht]
\centering
\includegraphics[width = 1.00\columnwidth]{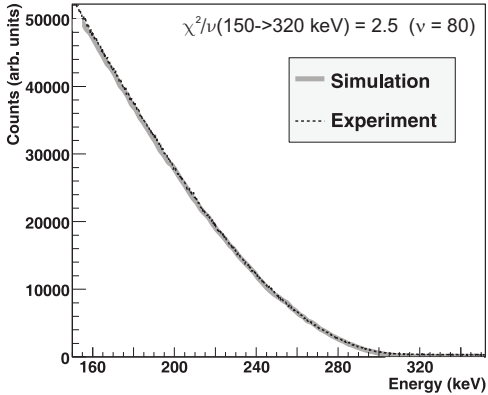}
\caption{Comparison between an experimental and a simulated spectrum for the $warm$ 13T data.} \label{fig:ExperimentVersusSimulation}
\end{figure}

The GEANT4 based Monte-Carlo simulation code, which was especially developed for this type of experiment \cite{WautersGeant2009}, dealt with the influence of the magnetic field on the trajectories of the electrons, the energy loss on their way to the detector and (back)scattering in the foil, on the sample holder and on the detector. To check and tune the quality of this simulation code a series of test measurements were performed under well controlled experimental conditions and then compared to the simulations. In the end, the $^{60}$Co $\beta$~spectrum could be reproduced with good precision in the region of interest (see Fig.~\ref{fig:ExperimentVersusSimulation}). Details of the testing and performance of this code are reported elsewhere \cite{WautersGeant2009}. All simulations used for the analysis of the experiment discussed here were performed with GEANT4.8.2 using the low-energy package \cite{webref:lowenergy} with the \emph{Cut For Secondairies} (CFS) parameter set to 10 $\mu$m and the $f_r$ parameter, which determines the step size for tracking of electrons at the boundary between two materials, set to 0.02 (see also Ref.~\cite{Kadri2007}).

As mentioned before, the complete experiment was simulated, and results were obtained by a comparison of the simulations with the experimental data via Eq.~(\ref{eq:Extract A}).
Accurate descriptions of the geometry of the source, detector, and surrounding materials were implemented in the simulation code. A detailed description of the detector was provided by the manufacturer. The depth distribution profile of the activity in the Cu foil was calculated on the basis of Fick's second law using the diffusion constant for Co in Cu from Ref.~\cite{Mackliet1958}. A fine gridded field map (resolution 1 mm) obtained from the manufacturer of the superconducting magnet provided the magnetic field at all positions.

To simulate the $cold$ data the angular distribution of the $\beta$~radiation, Eq.~(\ref{eq:W}), was implemented taking into account the values for the fraction $f$ and the temperature of the sample obtained from the $\gamma$~anisotropies, as well as the correct $v/c$ value of every emitted particle.

\begin{figure}[!ht]
\centering
\begin{tabular}{c}
\epsfig{file=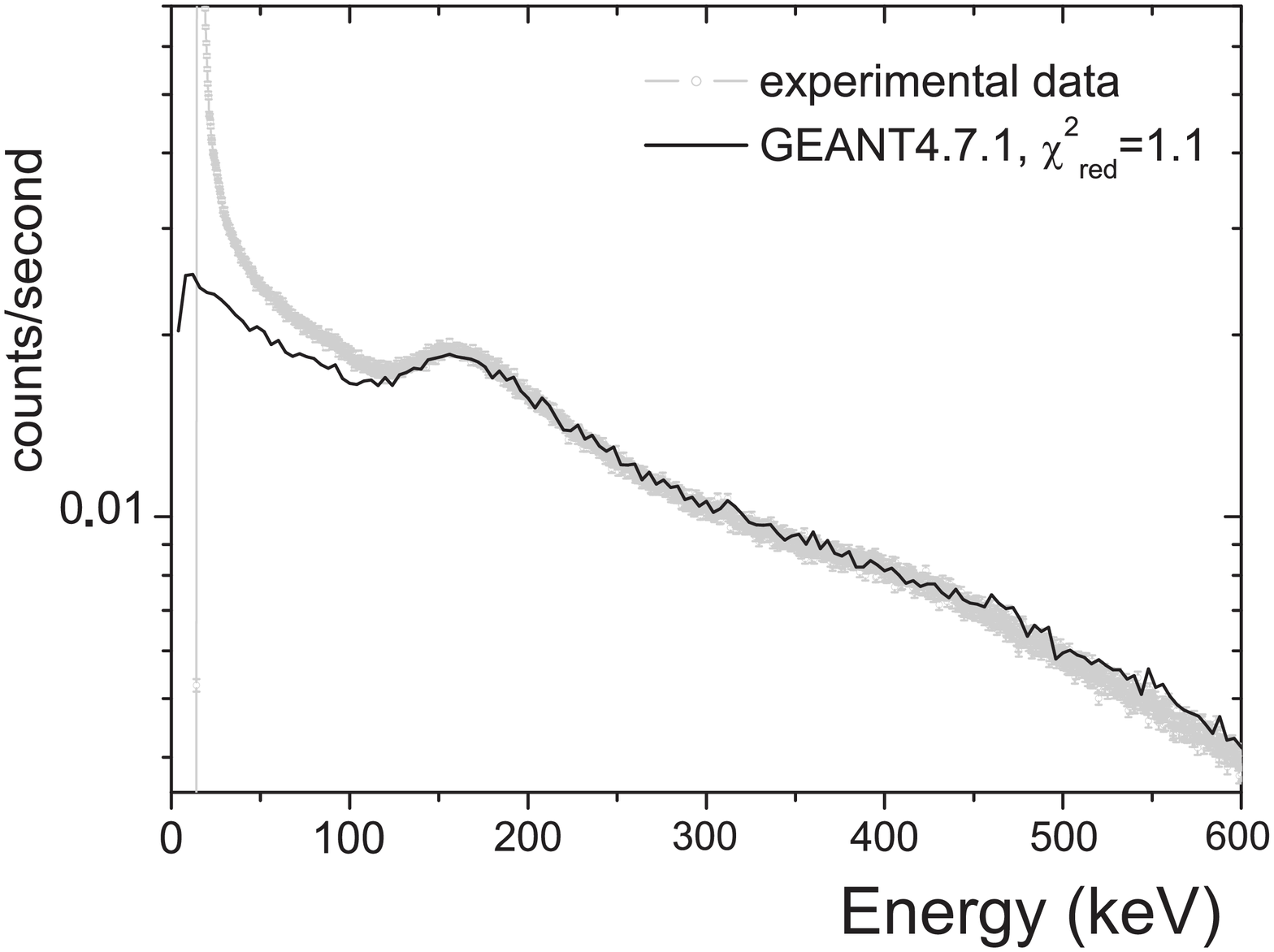,width=0.95\columnwidth}
\\[0.2cm]
\epsfxsize=7.5cm
\epsfig{file=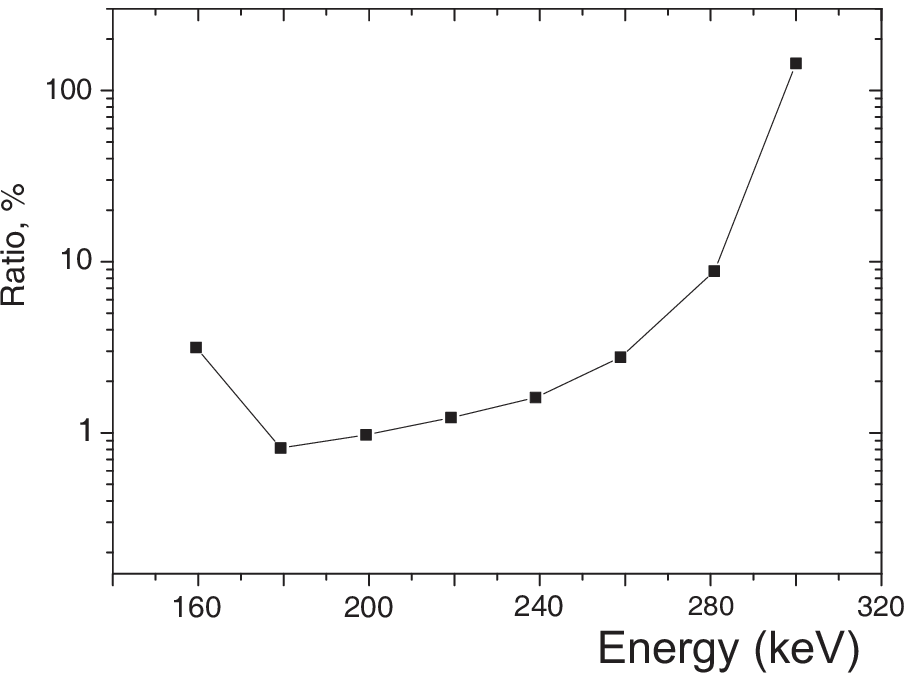,width=0.95\columnwidth}\\
\end{tabular}
\caption{Upper panel: Comparison of the experimental and simulated Compton background from the $\gamma$~rays
of $^{60}$Co. The experimental spectrum was obtained by installing a 0.4 mm thick copper plate in front of the detector to stop the $\beta$ particles. The quoted $\chi^2 / \nu$ comes from the difference between simulation and experiment in the region from 150 to 300~keV. Lower panel: Contribution of the Compton background in the upper half of the $\beta$~spectrum.}
\label{fig:Compton_BG}
\end{figure}

When processing the simulated data an energy resolution of 6~keV (see Sect.~II.B) and a  pile-up probability of $0.18~\%$ (obtained from the amplifier shaping time and the observed count rates) were applied.
As can be seen from Fig.~\ref{fig:ExpBetaSpectrum}, in order to obtain a pure $\beta$~spectrum, subtraction of the Compton background originating from the $\gamma$~rays of $^{60}$Co is essential.
Previous measurements showed that the shape of the Compton background could be reproduced very well by the simulations (Fig.~\ref{fig:Compton_BG}). The Compton background was therefore subtracted from both the simulated and experimental spectra using this shape and normalizing the intensity to the one observed in the region above the $\beta$~endpoint.
\\
The relative contribution from the Compton background to the total count rate in the energy region from 150~keV to 300~keV is shown in Fig.~\ref{fig:Compton_BG}.
In order to avoid large systematic effects due to this subtraction of the Compton background, two energy bins were excluded from the analysis. The last bin (from 290~keV to 320~keV) was excluded because there, the yield of the Compton electrons was comparable to that of the $\beta$~particles (see Fig.~\ref{fig:Compton_BG}). The first bin (from 150~keV to 170~keV) was excluded as well, as the Compton background in that energy region was not a smooth function of energy (Fig.~\ref{fig:Compton_BG}), so that any result from this bin would be too sensitive to the subtraction procedure. The Compton background yield for the bins remaining, bins 2-7, used for the final analysis, was of the order of a few percent of the total yield.
\\
\\

\subsection{Results and error analysis}

\begin{figure}[ht]
\centering
\includegraphics[width = 1.00\columnwidth]{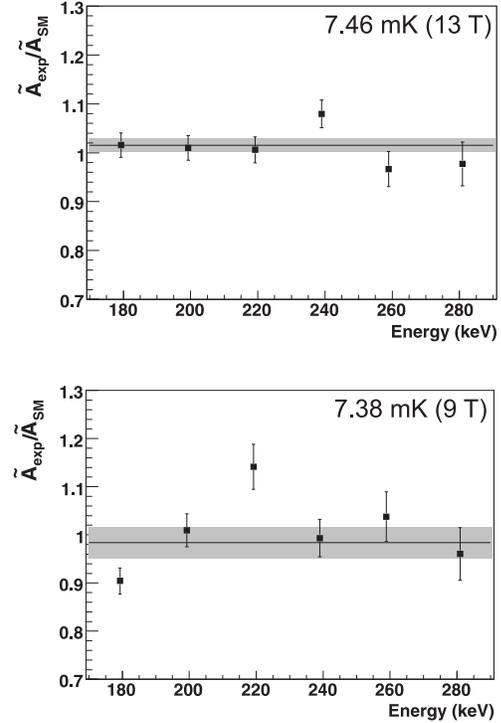}
\caption{Ratio between experimental and Standard-Model values for $\widetilde{A}$ obtained with Eq.~(\ref{eq:Extract A}) for the two measurements performed at the lowest temperatures. The grey band indicates the one standard deviation error on the weighted average.} \label{fig:SetsofA}
\end{figure}

The results for $\widetilde{A}$ as a function of energy are shown in Fig.~\ref{fig:SetsofA} for the two measurements performed at the lowest temperatures ({\it i.e.} largest degree of polarization). Within the current precision, no energy dependence was observed, as it should be. This means all effects that modify the size of the anisotropy in an energy dependent way, are all being dealt with correctly by the GEANT4 based Monte-Carlo code at this level of precision, which illustrates the power of our GEANT4 simulation-based analysis procedure. Further in this paper, it will be shown that the energy-dependent effect of the recoil terms on $\widetilde{A}$ is negligible at the current level of precision.

In Table~\ref{tab:Statistics} the results for $\widetilde{A}$ for each of the four measurements performed are listed with their statistical error bars.
These statistical errors are a combination of experimental and simulation statistics. If the spread of the individual data points was too large, the error bar on the weighted average of that data set was increased with a factor $\sqrt{\chi^2/\nu}$. Applying this procedure, the statistical error bar was increased on average by a factor of 1.6. Combining the four sets of data yields $\widetilde{A}_{exp} = -1.014(12)_{stat}$.
\\
\begin{table}[h]
\caption{\label{tab:Statistics}Values for $\widetilde{A}$ for each of the four measurements with their statistical error bars.}
\begin{ruledtabular}
\begin{tabular}{cccc}
    $B_{ext}$ (T)  & $T$ (mK) &  $\widetilde{A}$  & Stat. error\\
    \hline
    13  & 7.46 &  $-1.015$     & 15\\
    13  & 10.42  &  $-0.999$   & 45\\
    13  & 17.08  & $ -1.050$   & 32\\
    9  & 7.38 &  $-0.984$      & 32\\
    \hline
    Weighted average  & & &    $-1.014(12)$ \\
\end{tabular}
\end{ruledtabular}
\end{table}

A careful analysis of systematic uncertainties is crucial in this kind of precision experiments. The general approach was to vary a single parameter by one standard deviation and calculate the effect of this variation on the final result. If this was not possible using error propagation techniques, additional simulations with altered input were performed. The different contributions to the systematic error are listed in table \ref{tab:Systematics}. Below, every contribution will be discussed in more detail.
\begin{table}
\caption{\label{tab:Systematics} Different contributions to the systematic uncertainty on $\widetilde{A}$. The errors related to the GEANT4 simulations (*) are regarded as fully correlated and are therefore added linearly. The errors marked with ** are partially correlated and added accordingly.}

\begin{ruledtabular}
\begin{tabular}{lc}

    Effect  & Error (\permil) \\
    \hline
    Choice of region for analysis & 4 \\
    Pile-up correction  & 3 \\
    Background subtraction* & 2.5 \\
    Mismatch between simulation & \\ and experiment* & 3 \\
    Choice of GEANT4 physics parameters* & 5 \\
    Distribution of Co activity in Cu foil & 9 \\
    $\mu$B of $^{60}$Co(Cu) ** & 2 \\
    $\mu$B of $^{60}$Co(Cu) and $^{57}$Co(Cu) via & \\ the fraction $f_{\rm Co}$** & 3.2\\
    Fraction $f_{\rm Co}$ determination of Co(Cu)  & 4.3 \\
    Temperature determination with $^{54}$Mn & 3.2 \\
    Geometry of the setup & 0.2 \\
    \hline
    Total & 16 \\
\end{tabular}
\end{ruledtabular}
\end{table}

\subsubsection{Choice of energy region}

Selecting a specific part of the spectrum for the analysis introduces a bias. Changing the boundaries of the selected region ({\it i.e.} the region from 170~keV to 290~keV) by either including each of the outermost bins or not, and repeating the analysis showed that this bias leads to a 4~$\permil$ systematic error on $\widetilde{A}$.

\subsubsection{Pile-up}

The pile-up probability was determined to be $0.18~\%$ from the shaping time of the amplifier and the observed count rate. Trying to get this probability by other means ({\it e.g.} by integrating the pile-up induced tail at the righthand side of the pulser peak, Fig.~\ref{fig:ExpBetaSpectrum}), a $50~\%$ error bar to this probability was assigned, which leads to a $3~\permil$  error on the final result.

\subsubsection{Quality of the GEANT4 simulations}

The systematic errors that depend on the quality of the Monte-Carlo simulations are heavily correlated and were therefore added linearly.

Before the simulated and experimental data could be combined to extract $\widetilde{A}$, the Compton background had to be subtracted from the spectra. As this subtraction is not perfect \cite{WautersGeant2009}, a systematic error had to be assigned. By varying the shape of the Compton background ({\it i.e.} supposing a linear shape instead of the one shown in Fig.~\ref{fig:Compton_BG}, and varying the energy region above the $\beta$~spectrum endpoint that was used to determine the amplitude of the Compton background), the value of this error was estimated to be $2.5~\permil$.

A small difference between the experimental and simulated isotropic ($warm$) spectrum was observed, leading to a $\chi^2/\nu$ larger than unity (on average 2.6). This was translated into a systematic error on the integrals obtained from the simulated $\beta$~spectra. Propagating this error throughout the analysis yielded the corresponding systematic error on $\widetilde{A}$ of $3 \permil$.

Finally, two GEANT4 physics parameters were optimized for this particular application ({\it i.e.}~$f_r$ and CFS, see Sec. III.B) \cite{WautersGeant2009}. No obvious difference in the quality of the simulated spectra was seen between CFS = 1 $\mu$m and CFS = 10 $\mu$m. All simulations were thus performed with CFS = 10 $\mu$m because of the significantly shorter simulation time. The bias introduced by this choice was investigated by repeating one set of simulations with CFS = 1 $\mu$m, yielding a $5~\permil$ systematic error.

\subsubsection{Diffusion profile}

The amount of scattering and energy loss of the $\beta$~particles in the Cu foil depends in part on the depth profile of the $^{60}$Co activity in the foil. The depth profile resulting from diffusing the activity in the foil was calculated for the applied diffusion time of 300~s using the diffusion properties for Co in Cu reported in Ref.~\cite{Mackliet1958}. As the exact value of the diffusion time depends on the speed of heating and cooling of the oven we attached a generous error of 100~s to the value of 300~s. Reanalyzing then the set of data, this time using as input for the simulations the diffusion profile that was obtained by changing the diffusion time by 100~s, resulted in a shift of $\widetilde{A}$ of $9~\permil$. This was subsequently used as a systematic error related to the diffusion profile of the Co activity in the foil.

\subsubsection{Hyperfine interaction $\mu B$}

The error on the hyperfine interaction  strength $\mu B$ experienced by the $^{60}$Co nuclei ({\it i.e.} 51.83(16)~$\mu_{\text{N}}$T in a 13 T external field) induces an error on the nuclear polarization $P$, that is linear with $\widetilde{A}$, Eq.~(\ref{eq:W}). In addition, this error on $\mu B$ for $^{60}$Co, together with the error on $\mu B$ for the $^{57}$Co nuclei ({\it i.e.} 64.55(20)~$\mu_{\text{N}}$T in a 13 T external field; $\mu(^{57}$Co$) = +4.720(10) \mu_{\text{N}}$ \cite{Stone2005}), induces an error on $\widetilde{A}$ via the determination of the fraction~$f$. The error contributions to $\widetilde{A}$ caused by the uncertainties on $f$ and $P$ are correlated via $\mu B$ of the $^{60}$Co nuclei and were added accordingly.

\subsubsection{Fraction f}

As was discussed already, only a fraction $f$ = 0.928(4) of the $^{60}$Co nuclei were polarized. This fraction was determined from the anisotropies of four $\gamma$~lines in the decay of $^{60}$Co and $^{57}$Co. The measurement statistics, together with the error bar for the M1/E2 multipole mixing ratio of the 122~keV transition in the decay of $^{57}$Co, led to a $4.3~\permil$ error on the value of $f$, immediately leading to the same error on $\widetilde{A}$.

\subsubsection{Temperature determination}

The degree of polarization $P$ of the $^{60}$Co nuclei depends directly on the temperature of the Cu sample, which is obtained from the anisotropy of the 835~keV $\gamma$~ray of the $^{54}$Mn(Ni) nuclear thermometer. The error on this temperature depends on the amount of statistics, the precision to which the hyperfine interaction $\mu B$ of $^{54}$Mn(Ni) is known ({\it i.e.} $-64.20(17)$~$\mu_{\text{N}}$T in a 13 T external field), the fraction $f_{\rm Mn}$ of Mn nuclei that feel the full orienting hyperfine interaction in the Ni host (a calibration against a $^{60}$Co(Co) single crystal nuclear thermometer yielded $f_{\rm Mn}$ = 0.976(5)), and the accuracy to which the  position of the $^{54}$Mn source and the HPGe detector are known (negligible error contribution). All these effects together gave rise to a $3.2~\permil$ systematic error on $\widetilde{A}$ related to the temperature determination.

\subsubsection{Geometry}
Because of the large distances between the source and the two detectors, {\it i.e.}~19.5~cm ($\beta$~detector) and 30~cm ($\gamma$~detector), the accuracy to which the geometry of the setup was known affected the value of $\widetilde{A}$ only at the 2~x~$10^{-4}$ level.
\vspace{\baselineskip}
\\
Finally, on adding all errors in the correct way leads to our result $\widetilde{A}_{exp}$ = $-1.014(12)_{stat}(16)_{syst}$ with the main systematical errors originating from the GEANT4 simulations and from the depth profile of the activity in the sample foil (Table \ref{tab:Systematics}).
\\

\section{Recoil correction}

\subsection{Formalism}

Before we can interpret our result in terms of possible physics beyond the Standard Model, the value of $A_{SM}$ has first to be known with sufficient precision. This requires one to take into account the recoil corrections related to the induced weak currents, which occur because the decaying quark couples to the weak field as a bound particle in the nucleon, and not as a free particle. In particular, $^{60}$Co requires special attention as the $\beta$~decay is hindered by nuclear-structure effects leading to a rather high $\log ft$ of 7.5. As the allowed Gamow-Teller matrix element $M_{GT}$ is consequently strongly reduced, higher-order matrix elements become important and have to be considered.

The structure of the vector and axial-vector hadronic current for a $J = 1/2$ to $J' = 1/2$ $\beta$~decay has the following form \cite{Goldberger1958,Weinberg1958,Fuji1958}:
\begin{align}
V_{\mu}^{h} &=\bar{p}[g_{V}(q^{2})\gamma_{\mu}+g_{M}(q^{2})\sigma_{\mu\nu}\frac{q_{\nu}}{2M} \nonumber
\\&+ig_{S}(q^{2})\frac{q_{\mu}}{m}]n \nonumber \\ A_{\mu}^{h}&=\bar{p}[g_{A}(q^{2})\gamma_{\mu}\gamma_{5}+g_{T}(q^{2})\sigma_{\mu\nu}\gamma_{5}\frac{q_{\nu}}{2M} \nonumber
\\ &+ig_{P}(q^{2})\frac{q_{\mu}}{m}\gamma_{5}]n
\label{vgl:V+Arecoil}
\end{align}
with $q_\mu$ the four-momentum transfer, $M$ and $m$ the nucleon and the electron mass, respectively; $\sigma_{\mu \nu}$ and $\gamma_{\mu}$ are (combinations of) Dirac gamma matrices, and $p$ and $n$ the proton and neutron spinors. Further, $g_V$ and $g_A$ are the vector and axial-vector coupling constants, and $g_i$ ($i=M,S,T,P$) are the coupling constants of the weak-magnetism, scalar, tensor and pseudo-scalar induced weak currents, respectively. This form was generalized by Holstein \cite{Holstein1974}, who encoded all the nuclear-structure aspects of the problem into a few form factors denoted, $b$, $c$, $d$, $f$, $g$, $h$, $j_k$ with $k = 2, 3$.  These form factors depend on the coupling constants given in Eq.~(\ref{vgl:V+Arecoil}) and are functions of the momentum transfer $q^2$.  The Standard-Model value of the $\beta$-asymmetry parameter $\widetilde{A}$ can be expressed in terms of these form factors via two spectral functions \cite{Holstein1974}:
\begin{equation}
A_{SM}=\frac{H_{1}(E,J,J',0)}{H_{0}(E,J,J',0)} , \label{vgl:corrspectraalf}
\end{equation}
To first order in $1/M$ and with the positive sign convention for $g_A$ as adopted by \cite{Holstein1974}, this can be written, for a pure GT transition and with the upper(lower) sign for $\beta^-$($\beta^+$) decay, as:
\begin{align}
A_{SM,GT}^{\beta^{\mp}}=\mp\frac{\gamma_{JJ'}}{J+1} \Big[ 1
&+\frac{1}{A} \left ( \frac{E_e^2+2m^{2}}{3ME_e} \right ) \nonumber\\
\pm\frac{b}{Ac_{1}} \left ( \frac{E_e^2+2m^{2}}{3ME_e} \right )
&+\frac{d}{Ac_{1}} \left ( \frac{-E_e^2+m^{2}}{3ME_e} \right ) \nonumber\\
&\pm\frac{f}{Ac_{1}} \left ( \frac{\lambda_{JJ'}}{\gamma_{JJ'}}\frac{5E_e}{M} \right ) \Big]
\label{vgl:ASM}
\end{align}
with $A$ the mass number, $J$ and $J^{\prime}$ the spins of the mother and daughter nuclei, respectively, and
\begin{equation}
  \gamma_{J'J} =
\left\{ \begin{gathered}
  J+1 \;\text{for }J' = J - 1 \hfill \nonumber\\
  1 \;\text{for }J' = J \hfill \nonumber\\
  -J  \;\text{for }J' = J + 1 \hfill \nonumber\\
\end{gathered}  \right.
\label{vgl:gammaJJ}
\end{equation}
\begin{equation}
  \lambda_{J'J} =
\left\{ \begin{gathered}
  \sqrt{3(J-1)(J+1)}\;\text{for }J' = J - 1 \hfill\nonumber \\
  \sqrt{(2J-1)(2J+3)}\;\text{for }J' = J \hfill \nonumber\\
  \sqrt{3J(J+2)}\;\text{for }J' = J + 1 \hfill .\nonumber\\
\end{gathered}  \right.
\label{vgl:lambdaJJ}
\end{equation}

\noindent
Further, $c_1$ $=g_A M_{GT}$ is the leading Gamow-Teller form factor, which is the zero-momentum transfer limit of the $c = c_1 + c_2 q^2/M^2$ form factor, and $b$ and $d$ are the weak-magnetism and induced-tensor form factors, respectively. The $f$ form factor is less common and does not appear in the spectral functions for the $\beta$~spectrum shape and the $\beta - \nu$ correlation.  It is present in the expression for the $\beta$~asymmetry, $A_{SM}$, but its value is zero for $0^\pi \rightarrow 0^\pi$,  $0^\pi \leftrightarrow 1^\pi$ and $1/2^\pi \rightarrow 1/2^\pi$ transitions. The order $(1/M^2)$ form factors implicit in Eq.~(\ref{vgl:corrspectraalf}) but not explicitly shown in Eq.~(\ref{vgl:ASM}) are $g$, $h$, $c_2$ and $j_2$. Note that for the induced form factors, only the $q^2 = 0$ dependent term is considered.
\\

\subsection{Calculation of the induced form factors}

\begin{table*}[!hbtp]
\caption{\label{tab:recoil}Standard-Model $\beta$-asymmetry parameter, $A_{SM}$, including recoil corrections obtained with shell-model computed form factors from three different effective interactions. The correction was calculated for an average $\beta$ particle energy of 200~keV. Note that $A$ here stands for the atomic mass number. The sign conventions of Refs.~\cite{Holstein1974,Calaprice1977,Calaprice1976}, {\it i.e.} a positive $g_A$, was used.}
\begin{ruledtabular}
\begin{tabular}{cccccccc}
    Interaction & $\mid c_{1,exp} \mid$ & $b/Ac_1$ & $d/Ac_1$ & $f/Ac_1$ & $j_2/Ac_1$ & $g/Ac_1$  & $A_{SM}$\\
    \hline
    KB3     & 0.0138   & $-7.6$ & 4.4   & $-5.0$   & $-4.4 \times 10^5$   &  $5.6 \times 10^5$    & $-0.9779$ \\
    FPMI3   & 0.0138   & $-6.8$ & 3.4   & $-5.0$   & $-4.6 \times 10^5$   &  $5.6 \times 10^5$    & $-0.9767$ \\
    GXPF1A  & 0.0138   & $-6.4$ & $-4.3$  & $-3.1$   & $-3.0 \times 10^5$   &  $3.5 \times 10^5$    & $-0.9868$ \\
\end{tabular}
\end{ruledtabular}
\end{table*}

In light nuclei, the analogous $M1$ $\gamma$~transition to the Gamow-Teller transition can be measured.  This information, together with the conserved vector current (CVC) hypothesis \cite{Feynman1958}, enables the weak-magnetism form factor, $b$, to be determined from experimental data (e.g. \cite{Calaprice1976}).
In $^{60}$Co, however, no such information is available.  So all recoil-order form factors have to be calculated and for this we appeal to the nuclear shell model.
Our approach is that used by Calaprice, Chung and Wildenthal \cite{Calaprice1977}, where the form factors are expressed in terms of reduced matrix elements of standard spherical tensor operators.
To compute these reduced matrix elements for $^{60}$Co our preference would have been to perform an untruncated shell-model calculation in the full $fp$ shell.
This, however, was not practicable.
We settled for the model space, $f^{-1}r^5 + f^{-2}r^6$, where $f$ stands for the $f_{7/2}$ orbital and $r$ for any of the $f_{5/2}$, $p_{3/2}$, $p_{1/2}$ orbitals.  The configurations are expressed relative to a $^{56}$Ni closed-shell core.
Although this is a severe truncation, we note we are only computing recoil-correction terms for which great accuracy is not required.  Further, we will assign a generous error to the shell-model result, based on the spread of the values obtained with three different effective interactions:
a modified Kuo-Brown, KB3 interaction \cite{Poves1981,Poves1979}, the FPMI3 interaction \cite{Richter1991} and the GXPF1A \cite{Honma2002,Honma2005} interaction. Because of the truncated model space, the input single-particle energies were changed to reproduce the energies of the single-particle states in $^{57}$Ni.

Each of the three shell-model calculations produced very small Gamow-Teller matrix elements for the $5^+ \rightarrow 4^+$ transition in the decay of $^{60}$Co, as expected.  The experimental $\log ft$ value of 7.5 is a very large indicative of a small Gamow-Teller matrix element.  However, the particular shell-model value depends on subtle cancelations and may not be reliable.  Thus it was decided to fix the $c_1$ form factor, $c_1 = g_A M_{GT}$, to the experimental $ft$ value using
\begin{equation}
(ft)_{^{60}Co} \cong \frac{2  \mathcal{F}t^{0^+ \rightarrow 0^+}}{c^2_{1,exp}} ,
\label{eq:c_from_ft}
\end{equation}
with $\log(ft)_{^{60}Co}$ = 7.512(2) \cite{NNDC}, and $\mathcal{F}t^{0^+ \rightarrow 0^+}$ = 3072.08(79) s the corrected $ft$-value for the superallowed $0^+ \rightarrow 0^+$ transitions {\cite{Hardy2009}}. The value of $g_A$ was fixed to an effective value of $g_{A,eff} \cong 1$, typically used in finite nuclei \cite{Martinez:1996,Siiskonen:2001}. The same approach was adopted earlier to address the recoil corrections for $^{22}$Na \cite{Firestone1978}.

To calculate the effect of the induced form factors on $\widetilde{A}$ the full spectral functions $H_0$ and $H_1$ were used, including a Coulomb correction as given by Eq.~C4 in Ref.~\cite{Holstein1974}
\begin{align}
A_{SM} = & \left[ H_{1}(E,J,J',0)+\Delta F_1(E,J,J',0) \right] &\nonumber  \\[-0.9em]
&\line(1,0){150}& \nonumber \\[-0.5em]
&  [ H_{0}(E,J,J',0)+\Delta F_4(E,J,J',0)  &\nonumber \\
& +\frac{ \Delta F_7(E,J,J',0)}{3} ] .&
 \label{vgl:corrspectraalf_coulombcorr}
\end{align}

\begin{figure}[ht]
\centering
\includegraphics[width = 0.96\columnwidth]{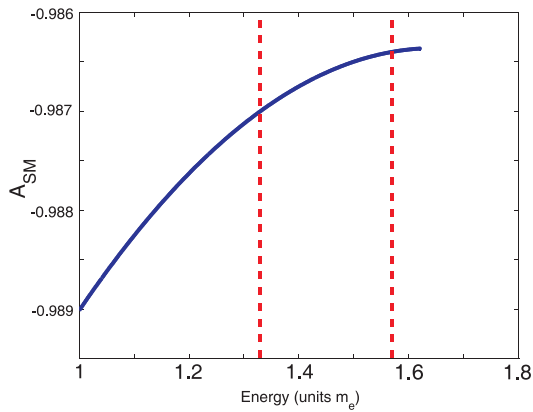}
\caption {The Standard-Model value of $\widetilde{A}$ including recoil corrections (GXPF1A calculation) versus electron energy. The dashed lines indicate the energy region used for analysis.} \label{fig:RecoilvsE}
\end{figure}

The results for the three effective interactions at an average electron energy of 200~keV are listed in Table~\ref{tab:recoil}. Although the effect of the recoil terms is energy dependent (Eq.~(\ref{vgl:ASM})) this dependence only plays a role below the $10^{-3}$ level in the energy region used for analysis (Fig.~\ref{fig:RecoilvsE}), and can therefore be neglected at the present level of precision. To interpret our experimental result in terms of tensor weak currents, the Standard-Model $\beta$-asymmetry parameter, including recoil corrections, as obtained with the GXPF1A effective interaction will be used, i.e. $A_{SM}$~=~-0.9868 (Table~\ref{tab:recoil}). This interaction is preferred as it was fitted to experimental data of nuclei in the mass range of $^{60}$Co (44 $< A <$ 67) \cite{Honma2002}, and has shown good performance for higher $fp$ shell nuclei, e.g. by reproducing the magnetic moment and the quadrupole moment of $^{60}$Co to within $10~\%$ of their experimental values (Table V and VI in Ref.~\cite{Honma2004}). The calculations using the KB3 and the FPMI3 interactions will be used to estimate the error on the recoil correction.
\\

\subsection{Relative importance of the different terms}

In cases where the Gamow-Teller matrix element is not hindered, the most important recoil form factors for $\beta$-asymmetry studies are $b$ and $d$, which occur in the combinations $b/A c_1$ and $d/A c_1$ (Eq.~\ref{vgl:corrspectraalf}).  For $^{60}$Co, however, where $c_1$ is small, one might expect these form factors to have a much greater importance.  But this turns out not to be the case: their influence on $A_{SM}$ does not exceed
a few permille (Table \ref{tab:effect_GXPF1A}). Moreover, the $b/Ac_1$ values (Table \ref{tab:recoil}) are not deviating too much from the values that were obtained for non-hindered transitions in mirror nuclei.  In these cases the $b$ form factor can be deduced from the isovector combination of magnetic moments \cite{Holstein1974} using the CVC hypothesis, with results that range from -2 to +8 \cite{Veronique2009,Calaprice1976}. Similar $b/Ac_1$ values for allowed but hindered decays were also obtained for the Gamow-Teller decays in the mass $A = 32$ isospin $T = 1$ triplet of $^{32}$P, $^{32}$S and $^{32}$Cl from the transition probability of the analog M1 $\gamma$~transition \cite{Holstein1974,Veronique2009}: for the $^{32}$P$ \rightarrow ^{32}$S $\beta^-$ decay ($\log ft = 7.9$) $b/Ac_1$ = 7.7 and for the $^{32}$Cl$ \rightarrow ^{32}$S $\beta^+$ decay ($\log ft$ = 6.7) $b/Ac_1$ = 2.0.

\begin{table}[!hbtp]
\caption{\label{tab:effect_GXPF1A} The effect on $A_{SM}$ of every induced form factor calculated with the GXPF1A effective interaction. Note that $f$ and $g$ are derived from the same quadrupole matrix element $M_Q$ \cite{Holstein1974}. The Coulomb correction included in Eq.~\ref{vgl:corrspectraalf_coulombcorr} has an effect of -0.10~\%.}
\begin{ruledtabular}
\begin{tabular}{cc}
    form factor & effect on $A_{SM}$ (\%) \\
    \hline
    $b$     & +0.33  \\
    $d$   & $-0.05$ \\
    $f$ and $g$ & +1.27   \\
    $h$ & 0.00  \\
    $j_2$ & $-0.13$ \\
    $c_2$ & 0.00 \\
\end{tabular}
\end{ruledtabular}
\end{table}
That the hindered $^{60}$Co decays would have $b/A c_1$ values comparable to
those of unhindered transitions can be understood as follows:
the $b$, $d$ and $c_1$ form factors can each be expressed as a matrix element of a rank-1 spherical tensor operator with an $M1$ character \cite{Calaprice1977,Holstein1974}. Thus if the $M_{GT}$ matrix element is suppressed, the other similar matrix elements are suppressed as well, leaving the ratios $b/A c_1$ and $d/A c_1$ to be of the same order as that for a fast transition.

However, when the rank-1 matrix elements are suppressed, the much less common $f$ and $g$ form factors become relatively more important.  These two form factors can be written in terms of matrix elements of a rank-2 spherical tensor with an $E2$ character \cite{Calaprice1977}. Such $E2$ matrix elements are not quenched, even though the $M1$ matrix elements are, resulting in ratios $f/A c_1$ and $g/A c_1$ being much larger than usual.  Indeed from Table~\ref{tab:effect_GXPF1A} it is seen the largest recoil correction to $A_{SM}$ comes from the $f$ and $g$ form factors, providing together roughly a $1~\%$ correction.

\subsection{Standard-Model value of $\mathbf{\widetilde{A}}$}

Before non-Standard-Model physics can be extracted from our experimental result, the error on the recoil corrected Standard-Model value of $\widetilde{A}$ has to be estimated. To do this, the difference between the form factors calculated with the GXPF1A effective interaction and the KB3 and FPMI3 interactions was taken as the error on each induced form factor. On noting the correlation between the $f$ and $g$ form factors, we finally recommend a Standard-Model value of
\begin{equation}
A_{SM} = -0.987(9) .
\label{A_SM}
\end{equation}
Being more conservative, one could also assume a $100~\%$ error on the total recoil correction, resulting in a Standard-Model value of $A_{SM} = -0.987(13)$.

\section{Discussion}

Our value for $\widetilde{A} = -1.014(20)$ is one of the most accurate measurements to date for a $\beta$-asymmetry parameter in a nuclear decay. It is in agreement with the Standard-Model prediction, corrected for the induced weak currents, Eq.~(\ref{A_SM}). Our result is comparable to the value previously obtained by Chirovsky {\it  et al.}, of $-1.01(2)$ \cite{Chirovsky1984}. When investigating the report on their measurement in more detail, we recognize these authors' efforts to reduce all sources of systematic errors. However, it is unclear whether the quoted error bar is purely statistical or whether systematic errors, such as the remaining scattering of the $\beta$~particles were also included. Indeed no detailed list of systematic errors was given, while in later works on $^{56}$Co and $^{58}$Co using the same setup systematic errors were addressed in detail and were of the order of a few percent \cite{Lee1983,Lee1985}. The effect of scattering was clearly demonstrated by these authors by presenting results for a $thin$ and a $thick$ source, which yielded values for the asymmetry parameter that differed by as much as $30~\%$. Our $^{60}$Co(Cu) sample is very similar to their \emph{thin} source, and the remaining systematic uncertainties in our measurement are listed in Table \ref{tab:Systematics}. Note that we used GEANT4 based Monte-Carlo simulations to address systematic effects caused by scattering and the magnetic field.

This is the first time the influence of the induced weak currents on the $\beta$-asymmetry parameter of $^{60}$Co was investigated. Shell-model calculations demonstrated that these can not be ignored if our experimental value, $\widetilde{A}_{exp} = -1.014(20)$, is to be interpreted in terms of non Standard-Model physics.

\subsection{Limits on tensor currents}

As a possible time-reversal violating tensor current is already strongly restricted by a measurement of the electron transverse polarization in the decay of $^8$Li \cite{Huber2003}, our result is mainly sensitive to the real, {\it i.e.} time-reversal invariant, terms in Eq.~(\ref{eq:asym}). To first order in $C_T$ and $C_T^{\prime}$ one has from the last line of Eq.~(\ref{eq:asym})
\begin{equation}
\widetilde{A}_{exp} = A_{SM} + \frac{\gamma m} {E_e} \Re \left(\frac{ C_T + C^{\prime}_T }{ C_A } \right) ,
\label{vgl:A_Ct_firstorder}
\end{equation}
\noindent with the sensitivity factor $\gamma m/E_e$ = 0.704 for the measurement reported here (the maximal value being unity). Our result is thus mainly sensitive to a time-reversal invariant but parity-violating left-handed tensor component in the weak interaction. From Eq.~(\ref{vgl:A_Ct_firstorder}) taking the Standard-Model prediction of Eq.~(\ref{A_SM}) we can then extract the limits: $-0.088 <  (C_T + C^{\prime}_T )/C_A < 0.014$ (90~~\% C.L.) for the amplitudes of the tensor coupling constants. If we adopt the more conservative $100~\%$ error on the recoil correction, these limits are not much affected, {\it i.e.} $-0.094 <  (C_T + C^{\prime}_T )/C_A < 0.018$ (90~~\% C.L.).

\begin{figure}[h]
\centering
\includegraphics[width = 1.00\columnwidth]{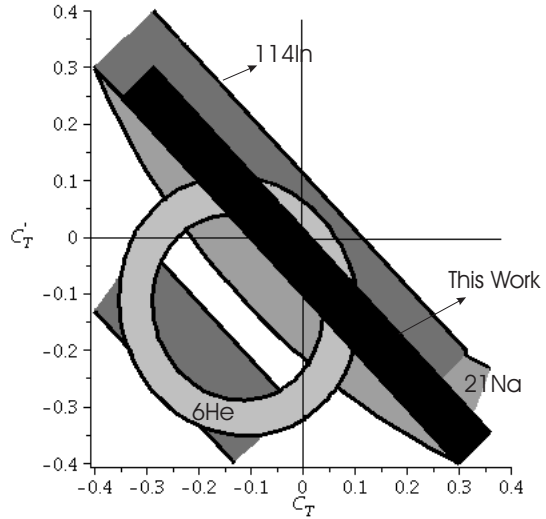}
\caption{Limits on $C_T$ and $C^{'}_{T}$ derived from our $\beta$ asymmetry parameter result for $^{60}$Co compared to other experiments in nuclear $\beta$~decay sensitive to a possible tensor contribution. Results are shown for the beta-neutrino correlation $a_{\beta\nu}$ for $^6$He \cite{Johnson:1963}, a measurement of the $\beta$-asymmetry parameter $A$ for $^{114}$In \cite{WautersIn2009} and a measurement of $a_{\beta\nu}$ for $^{21}$Na \cite{Vetter:2008}. The bands on the graph represent $90~\%$ C.L. intervals.}
\label{fig:LimitsCT}
\end{figure}

These limits are compared to limits from other precise experiments in Fig.~\ref{fig:LimitsCT}. As can be seen, our result provides limits that are competitive to the ones from the most sensitive experiment to date searching for time-reversal invariant tensor couplings in nuclear $\beta$~decay, {\it i.e.} the measurement of the $\beta-\nu$ correlation of $^6$He performed by Johnson et al. \cite{Johnson:1963}.
\noindent With our current experimental precision, the width of the band of allowed values for $C_T$ and $C_T^{\prime}$ in Fig.~\ref{fig:LimitsCT} is not much affected by the recoil corrections on $A_{SM}$. The position of the band, however, does depend on the accuracy of the shell-model calculations.

\subsection{Limits on right-handed currents}

\begin{figure}[h]
\centering
\includegraphics[width = 0.9\columnwidth]{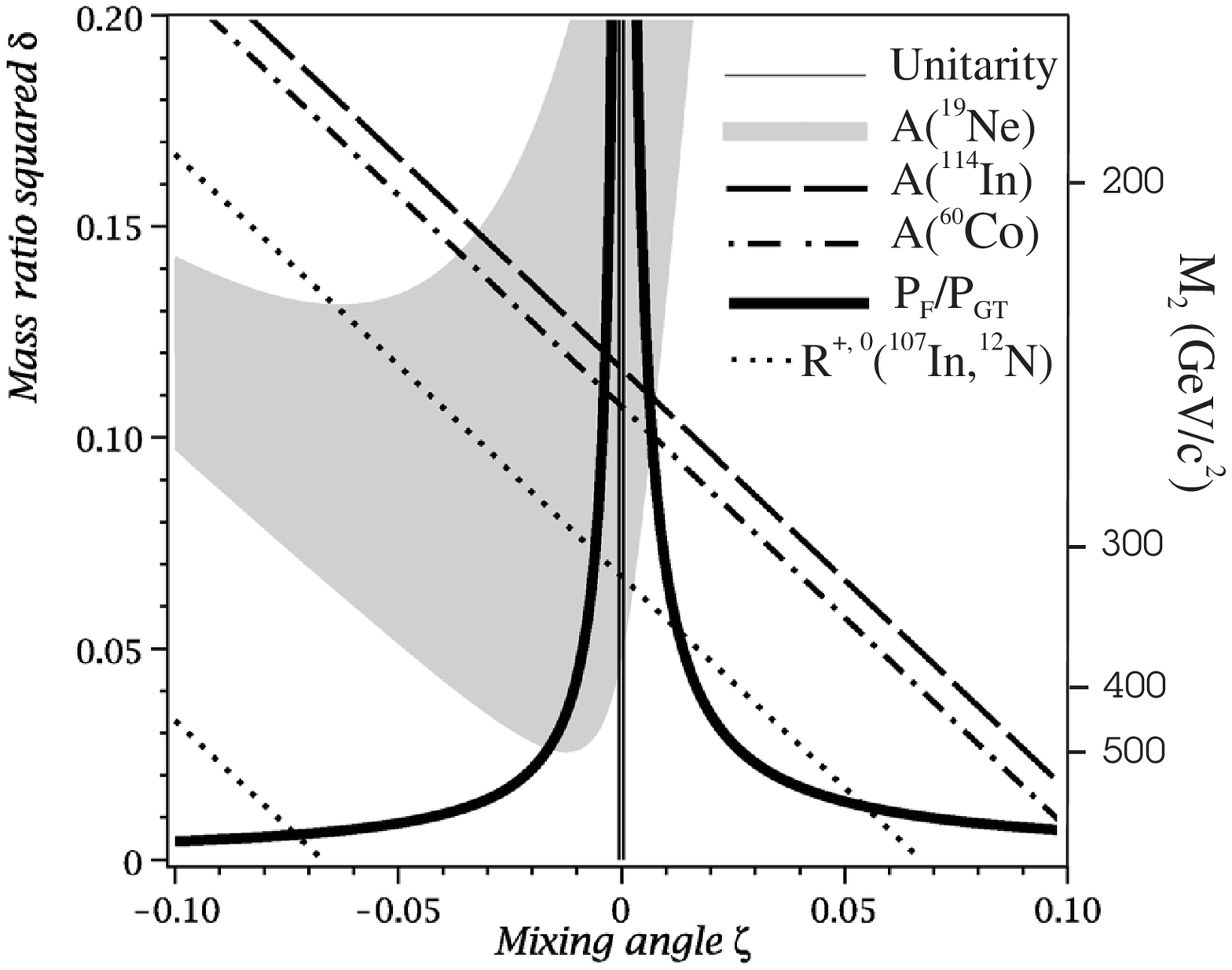}
\caption{Constraints on the right-handed currents parameters
$\delta$ and $\zeta$ from measurements of different observables in
nuclear $\beta$ decay, i.e. the test of unitarity of the quark mixing matrix \cite{Hardy2009}, measurements of the $\beta$ asymmetry parameter in the decays of $^{19}$Ne \cite{Calaprice1975}, $^{114}$In \cite{WautersIn2009} and $^{60}$Co (this work), relative measurements of the longitudinal polarization of $\beta$ particles in pure Fermi and Gamow-Teller transitions \cite{carnoy90,wichers87} and relative measurements of the longitudinal polarization of $\beta$ particles emitted by polarized nuclei \cite{severijns93,severijns98,allet96,thomas01}}
\label{fig:rhc}
\end{figure}

Our result can also be interpreted in terms of right-handed currents. For ease of comparison we limit here to the so-called manifest (or minimal) left-right symmetric (MLRS) model \cite{beg77}. This explains a departure from maximal parity violation entirely by the presence of a second W boson, $W_R$, which mediates right-handed couplings, in addition to the usual $W_L$ boson. These weak interaction eigenstates are linear combinations of the mass eigenstates $W_1$ and $W_2$ (with masses $m_1$ and $m_2$):
\begin{eqnarray}
W_L  =    W_1 cos \zeta + W_2 sin \zeta  \nonumber \\
W_R  =   -W_1 sin \zeta + W_2 cos \zeta
\end{eqnarray}
with $\zeta$ the mixing angle (a CP violating phase $\omega$ is neglected). The MLRS model assumes the coupling constants and Cabibbi-Kobayashi-Maskawa (CKM) quark mixing matrix elements for left-handed and right-handed couplings to be identical such that there are only two parameters, i.e. $\zeta$ and $\delta = (m_1/m_2)^2$, which are both zero in the standard model.
In this model $A$~=~$A_{SM}[1 - 2(\delta + \zeta)^2]$ for a pure Gamow-Teller transition as was studied here.
Figure~\ref{fig:rhc} shows the limits for the parameters $\delta$ and $\zeta$ from this and other experiments in nuclear $\beta$ decay. A similar graph can be made for experiments in neutron decay (see e.g. Ref.~\cite{schumann07a}) but, due to the mixed Fermi/Gamow-Teller character of neutron decay, the constraints in this case depend on the neutron lifetime, $\tau_n$, the value of which is at present unclear due to a large discrepancy between the two most precise values that are available to date \cite{arzumanov00,serebrov05,paul09}.
For $\zeta \simeq 0$, as follows from the unitarity condition for the CKM matrix \cite{Hardy2009}, our result corresponds to a lower limit of $M_2$ = 245 GeV/c$^2$ (90\% C.L.) for the mass of the weak boson eigenstate $W_2$ that is mainly related to a $W_R$ boson. This is to be compared to values ranging from 220 GeV/c$^2$ to 310 GeV/c$^2$ (90\% C.L.) obtained from other experiments in nuclear and neutron $\beta$ decay \cite{severijns93,severijns98,allet96,thomas01,serebrov98,schumann07,schumann07a}.

\subsection{Combined value for $\widetilde{A}$ for $^{60}$Co}

The combined value of our experimental $\widetilde{A}$ and the two other values available in the literature (see Table~\ref{tab:Abig}), {\it i.e.} $\widetilde{A}_{combined}$($^{60}$Co) = -1.006(13), would provide stringent limits on time-reversal invariant tensor currents and right-handed currents, and could possibly also contribute to the search for supersymmetric contributions \cite{profumo07} if more accurate values for the induced form factors would be available. Therefore, additional theoretical calculations and/or experimental efforts ({\it e.g.} spectroscopic information on the analog $\gamma$~ transition in the daughter nucleus $^{60}$Ni or a measurement of the $^{60}$Co $\beta$ spectrum shape) would be of great value.

Note, finally, that this combined value of the $\beta$-asymmetry parameter of $^{60}$Co also limits the size of higher-order corrections on $A_{SM}$ for this isotope to about $2~\%$ (90~\% C.L.).
\\

\section{Conclusion}

A new determination of the $\beta$-asymmetry parameter $\widetilde{A}$ in the $\beta$~decay of $^{60}$Co was reported. In the analysis extensive use was made of a newly developed GEANT4 based Monte-Carlo simulation code to include the effects of scattering and of the polarizing magnetic field. The size of the induced matrix elements was addressed for the first time for this isotope by performing shell-model calculations using three different effective interactions. Our experimental value of $\widetilde{A} = -1.014(20)$, one of the most accurate values for this parameter in a nuclear decay ever obtained, is in agreement with the recoil corrected Standard-Model value of $-0.987(9)$. It provides limits for the amplitudes of tensor coupling constants in the charged-current weak interaction that are competitive to those obtained in previous experiments and reduces the present upper limits on such coupling constants. Limits on right-handed currents were obtained as well, although these are somewhat less stringent than the ones from previous experiments.

Finally, the shell-model calculation of the induced form factors showed that for an allowed but hindered transition, such as the $\beta$~decay of $^{60}$Co, it is not the common weak-magnetism, $b$, and induced tensor, $d$, form factors that determine the size of the recoil corrections on $\widetilde{A}$, but rather the E2-type form factors $f$ and $g$.

This work was supported by the Fund for Scientific Research
Flanders (FWO), project GOA/2004/03 of the K. U. Leuven, the Interuniversity Attraction Poles Programme, Belgian
State Belgian Science Policy (BriX network P6/23), and the grant LA08015
of the Ministry of Education of the Czech Republic.

\end{document}